\documentclass[prd,
twocolumn,
superscriptaddress,nofootinbib,%
tightenlines,showpacs,showkeys]{revtex4}

\usepackage{amsfonts,amsmath,graphicx,epsfig,amssymb}
\usepackage{epsfig}
\usepackage{mathrsfs}
\usepackage{latexsym}
\usepackage{amsxtra}
\usepackage{amsbsy}
\usepackage{amscd}
\usepackage{color}
\usepackage{bbm}
\usepackage{hyperref}
\usepackage{multirow}

\usepackage{graphics}
\usepackage{amsfonts}

\usepackage{subfigure}
\usepackage{color}
\allowdisplaybreaks

\newcommand{\rbox}{\rule[-0.20cm]{0cm}{5mm}}

\newcommand{\be}{\begin{equation}}
\newcommand{\ee}{\end{equation}}

\def\no{\nonumber}
\def\bea{\arraycolsep .1em \begin{eqnarray}}
\def\eea{\end{eqnarray}}

\begin{document}
\title{Pole analysis on the hadron spectroscopy of $\Lambda_b\to J/\Psi p K^-$}
\author{Shi-Qing Kuang}
\affiliation{School of Physics and Electronics, Hunan University, Changsha 410082, China}
\author{Ling-Yun Dai}
\email{dailingyun@hnu.edu.cn}
\affiliation{School of Physics and Electronics, Hunan University, Changsha 410082, China}
\author{Xian-Wei Kang}
\affiliation{College of Nuclear Science and Technology, Beijing Normal University, Beijing 100875, China}
\author{De-Liang Yao}
\affiliation{School of Physics and Electronics, Hunan University, Changsha 410082, China}
\begin{abstract}
In this paper we study the $J/\Psi p$ spectroscopy in the process of $\Lambda_b\to J/\Psi p K^-$.
The final state interactions of coupled channel $J/\Psi p$ ~-~ $\bar{D} \Sigma_c$~-~$\bar{D}^{*} \Sigma_c$ are constructed based on K-matrix with the Chew-Mandelstam function.
We build the $\Lambda_b\to J/\Psi p K^-$ amplitude according to the Au-Morgan-Pennington method. The event shape is fitted and the decay width of $\Lambda_b\to J/\Psi p K^-$ is used to constrain the parameters, too.  With the amplitudes we extract out the poles and their residues.  Our amplitude and pole analysis suggest that the $P_c(4312)$ should be $\bar{D}\Sigma_c$ molecule, the $P_c(4440)$ could be an S-wave compact pentaquark state, and the structure around $P_c(4457)$ is caused by the cusp effect.
The future experimental measurement of the decays of $\Lambda_b\to \bar{D}\Sigma_c K^-$ and $\Lambda_b\to \bar{D}^*\Sigma_c K^-$ would further help to study the nature of these resonances.
\end{abstract}
\pacs{11.55.Fv, 11.80.Et, }
\keywords{Dispersion relations, Partial-wave analysis}

\maketitle

\parskip=2mm
\baselineskip=3.5mm

\section{Introduction}\label{sec:introduction}
The discovery of hidden-charm hadrons $P_c^+(4380)$ and  $P_c^+(4450)$ \cite{Aaij:2015tga} started a new era of hadron physics as they obviously contain at least five quark component $\bar{c}cuud$.
Whether they are compact pentaquark states or hadronic molecules or generated by kinematic effect is still not clear. For recent review on the hadronic molecules and multiquark states, we refer to \cite{Guo:2017jvc,Liu:2019zoy,Brambilla:2019esw}.
Recently the LHCb experiment made a great progress \cite{Aaij:2019vzc}. The decay events collected now by Run 1 and Run 2 are about nine times more than that of Run 1 analysis. As a result, the bin size has been decreased from 15 to 2~MeV.
With the high statistics they found three structures in the $J/\Psi p$ spectrum of $\Lambda_b\to J/\Psi p K^-$:
\bea
P_c^+(4312):\; M&=&4311.9\pm0.7^{+6.8}_{-0.6}\;,  \no\\
 \Gamma&=&9.8\pm2.7^{+3.7}_{-4.5} \;,  \no\\
P_c^+(4440):\; M&=&4440.3\pm1.3^{+4.1}_{-4.7}\;,  \no\\
 \Gamma&=&20.6\pm4.9^{+8.7}_{-10.1}\;,  \no\\
P_c^+(4457):\; M&=&4457.3\pm0.6^{+4.1}_{-1.7}\;,  \no\\
 \Gamma&=&6.4\pm2.0^{+5.7}_{-1.9}\;,  \no
\eea
with all the units being MeV.
Then the question is, what inner structure are they? A cornucopia of models have been done to study the property of these resonances  \cite{Skerbis:2018lew,Chen:2019bip,Liu:2019tjn,Guo:2019fdo,He:2019ify,
Guo:2019kdc,Fernandez-Ramirez:2019koa,Wang:2019got,Wang:2019nwt,
Cheng:2019obk,Wu:2019adv,Xiao:2019aya,Voloshin:2019aut,Xu:2019zme,Liu:2019zvb,Burns:2019iih,Du:2019pij}.
Among them, Ref.\cite{Fernandez-Ramirez:2019koa} uses a coupled channel K-matrix formalism to fit to the data around the $P_c(4312)$. They found
the attractive effect of the $\Sigma_c^+ \bar{D}^0$ channel, but it is not strong enough to form a bound state.
In Ref.\cite{Du:2019pij}, the Lippmann-Schwinger equations have been used and the one pion exchange and short range scattering potential have been considered.
Thay found the three resonances and also a narrow $\Sigma_c^*\bar{D}$ state ($P_c(4380)$). In addition, three more $\Sigma_c^*\bar{D}^*$ molecules have been seen in the analysis. All these poles are hadronic molecules of $\Sigma_c^{(*)}\bar{D}^{(*)}$ channels. Since lots of the paper support the molecule picture of these $P_c$ states,
it would be rather interesting to distinguish the molecule and non-molecule structure.
The pole counting rule\cite{Morgan:1992ge,Dai:2011bs} is a right way to do such study. Indeed the shadow poles (accompanying the ones being closest to the physical sheet) are also important to discuss the inner structure. Here we use the Chew-Mandelstam formalism to write the unitary cut in once subtracted dispersion relation, and we fit up to 4.6~GeV to discuss the three resonances listed above ($P_c(4312)$, $P_c(4440)$, and $P_c(4457)$). Our amplitudes describe the data around the $P_c(4440)$ and $P_c(4457)$ region well and they are helpful for discussing the structure. These will be discussed in next sections.

To extract the information of these resonances, we need amplitude analysis to describe the invariant mass
spectroscopy of $\Lambda_b\to J/\Psi p K^-$. The final state interactions (FSI) are important to be considered. There have been lots of papers that indicate the importance of the FSI, see e.g. \cite{Au:1986vs,Dai:2012pb,Kang:2013jaa,Dai:2014zta,Guo:2015dha,Chen:2016mjn,Dai:2017fwx,Danilkin:2019mhd,Cao:2019wwt}. We will use the Au-Morgan-Pennington (AMP) method \cite{Au:1986vs} to include the FSI of $J/\Psi p$ ~-~ $\bar{D} \Sigma_c$~-~$\bar{D}^{*} \Sigma_c$ triple channels.  From the amplitudes we extract out the pole information and discuss their property according to the pole counting rule.

This paper is organized as follows: In Sect.~II we use
K-matrix to build the hadronic scattering amplitudes of $J/\Psi p$ ~-~ $\bar{D} \Sigma_c$~-~$\bar{D}^{*} \Sigma_c$ triple channels. And the $\Lambda_b\to J/\Psi p K^-$ amplitude is constructed by the AMP method.
In Sect.~III we fit to the event shape and decay width of $\Lambda_b\to J/\Psi p K^-$ and determine the parameters.
The amplitudes are continued into un-physical Riemann sheets (RSs) and the poles in different RSs are extracted out.
By pole analysis the origin of these poles are discussed.
In Sect.~IV we discuss the fits to other datasets given by the LHCb.
We end with a brief summary.

\section{Decay amplitude}\label{sec:formalism}
To get the information of poles, we need an amplitude analysis to get accurate hadronic scattering amplitudes. The following problem is which channel should be included? As is predicted in \cite{Wu:2010jy,Wu:2010vk}, there could be $\bar{D}\Sigma_c$ and  $\bar{D}^{*}\Sigma_c$ hadronic molecule states with quantum number $IJ^P=\frac{1}{2}\frac{1}{2}^-$ at $4261+i28.5$ and $4412+i23.6$~MeV of each channel. This is later studied in \cite{Xiao:2013yca} by a coupled channel unitary approach.
In Ref.\cite{Aaij:2019vzc}, the $P_c^+(4312)$ is found to be under $\bar{D}\Sigma_c$ threshold and more like an S-wave resonance.  The other two resonances are proximate to the $\bar{D}^{*}\Sigma_c$ threshold. We thus take $J/\Psi p$ ~-~ $\bar{D} \Sigma_c$~-~$\bar{D}^{*} \Sigma_c$  as the coupled channels.
The helicity has been ignored as that the heavy quark spin symmetry ensures the
spin-dependent interactions related to the heavy quark are of the order of $1/m_Q$ \cite{Manohar:2000dt}\footnote{We are aware of that the spin dependent interactions between the light quarks can not be ignored, and we refer readers to read Ref.\cite{Du:2019pij}.}. These assumptions are consistent with the analysis of Ref.\cite{Aaij:2019vzc}, where it is also found that including P-wave factors in the Breit-Wigner amplitudes has negligible effect on the results.

In addition, the thresholds of $\Sigma^{*+}_c \bar{D}^{0}$ ($\Sigma^{*++}_c D^{-}$) and  $\Sigma^{*+}_c \bar{D}^{*0}$ ($\Sigma^{*++}_c D^{*-}$) are 4382.33 (4388.06)~MeV and 4524.35 (4528.67)~MeV, respectively. They are far away from the three resonances we studied here. Besides, from the LHCb experiment measurement\cite{Aaij:2019vzc}, there is no obvious structure around these thresholds, we thus do not consider their contribution here.  For the $\Lambda_c^+\bar{D}^{(*)0}$ channels, the mass and width of the $P_c(4312)$ are $M=4311.9\pm0.7^{+6.8}_{-0.6}$~MeV and $\Gamma=9.8\pm2.7^{+3.7}_{-4.5}$~MeV. Taking into account that the $\Lambda_c^+\bar{D}^{(*)0}$ threshold is 19~MeV below the peak of the $P_c(4312)$, twice as much as the width, and the $\Lambda_c^+\bar{D}^0$ is even lower, they won't contribute a lot. It should also be pointed out that their interactions are expected to be repulsive and can not form the hadronic bound state\cite{Aaij:2019vzc}. Thus all the channels, except for the $J/\Psi p$ ~-~ $\bar{D} \Sigma_c$~-~$\bar{D}^{*} \Sigma_c$, are not included in our model.

We construct our amplitude based on K-Matrix to keep unitarity, and have
\bea
T(s)=K(s)[1-C(s) K(s)]^{-1}\,,\label{Eq:KM;T}
\eea
where $\sqrt{s}$ is the energy in the center of mass frame. $K(s)$ is a real matrix and it could be parameterized as
\bea
K^{ij}(s)&=&\sum_{l}\frac{f^{i}_{l}f^{j}_{l}}{(s_{l}-s)}
+\sum_{n=0}c^{ij}_{n}(\frac{s}{s_{th1}}-1)^{n}\;.\label{eq;K}
\eea
To reduce the model dependence, we include as less as possible parameters. According to practice we set $c^{ij}_{n\geq2}=0$ and  $s_{l\geq2}=0$.
$C(s)$ is the diagonal matrix of the canonical definition of Chew-Mandelstam function \cite{Chew:1960iv,Edwards:1980sa}, and it could be written in once subtracted dispersion relation:
\bea
C_i(s)=\frac{s}{\pi}\int_{s_{thi}}^{\infty} ds'\frac{\rho_i(s')}{s'(s'-s)},\label{eq:C}
\eea
where $i(j)=1,2,3$ represent for $J/\Psi p$, $\bar{D}\Sigma_c$, $\bar{D}^{*} \Sigma_c$, respectively, with isospin $1/2$. $s_{thi}=(M_i+m_i)^2$ is the threshold and $M_i$ ($m_i$) is the mass of meson (baryon) in the $i$-th channel.
The phase space factor has only diagonal elements:
\be
\rho_i(s)=\sqrt{\frac{\left(s-(M_i+m_i)^2\right)\left(s-(M_i-m_i)^2\right)}{s^2}} \,.\label{eq;rho}
\ee
The Chew-Mandelstam function could be expressed explicitly as
\bea
C_i(s)&=&\frac{1}{\pi}+\frac{M_i^2-m_i^2}{\pi s}\ln\left(\frac{m_i}{M_i}\right)-\frac{M_i^2+m_i^2}{\pi (M_i^2-m_i^2)}\ln\left(\frac{m_i}{M_i}\right)   \no\\
      &+&\frac{\rho_i(s)}{\pi}\ln\left(\frac{\sqrt{(M_i+m_i)^2-s}-\sqrt{(M_i-m_i)^2-s}}{\sqrt{(M_i+m_i)^2-s}+\sqrt{(M_i-m_i)^2-s}}\right)\,.  \no\\
      \label{Eq:C;ana}
\eea

The new high statistics results of $J/\Psi p$ line shape (in the process of $\Lambda_b\to J/\Psi p K^-$) from LHCb \cite{Aaij:2019vzc} help us to constrain the hadronic scattering amplitude. To describe it, the rescattering of inelastic channels ($\bar{D}\Sigma_c$, $\bar{D}^*\Sigma_c$) needs to be considered. We implement the AMP formalism \cite{Au:1986vs} to include the FSI:
\bea
F_i(s)&=&\sum_{k=1}^3\alpha_k(s) T_{ki}(s) \; , \label{eq:F}
\eea
where '$i$' and '$k$' have the same meaning as explained after Eq.~(\ref{eq:C}). The $\alpha_k(s)$ are polynomials of $s$, absorbing all the contributions of left hand cut and distant right hand cut. For simplicity we set them to be constant $\alpha_{1,2,3}$ and ignore higher order terms. It is easy to check that Eq.~(\ref{eq:F}) satisfies the final state interactions theorem:
\bea
{\rm Im}F_i(s)=\sum_{k=1}^3 F_{k}^*\rho_k(s) T_{ki}(s) \, . \label{eq:unit}
\eea
With this amplitude we fit to the invariant mass spectroscopy
\bea
\frac{d\Gamma_i}{d\sqrt{s}}=\frac{ \lambda^{1/2}(s,M_{J/\Psi}^2,m_p^2)\lambda^{1/2}( M_{\Lambda_b^0}^2,s,m_K^2)|F_i|^2}{256\pi^3 M_{\Lambda_b^0}^3 \sqrt{s}}\;. \label{eq:dGds}
\eea
Here the K\"all\'en function $\lambda$ is defined as $\lambda(x,y,z)=(x-y-z)^2-4yz$.
To fit to the invariant mass spectroscopy one would need to time $\frac{d\Gamma_i}{d\sqrt{s}}$ with a normalization factor \lq $N$'.
The decay width given by the PDG \cite{Tanabashi:2018oca} could be used to constrain the $\alpha_i$, too. But still we lack adequate constraints on $F_2(s)$ and $F_3(s)$ amplitudes. Indeed in Ref.\cite{Xiao:2019aya}, the ratio of the coupling constants  $\frac{g_{P_c(4306)\bar{D}\Sigma_c}}{g_{P_c(4306)J/\Psi p}}$ and $\frac{g_{P_c(4453)\bar{D}^*\Sigma_c}}{g_{P_c(4453)J/\Psi p}}$ are roughly 4 and 2 times, while the ratios of the phase spaces (with $F_i(s)=1$ in Eq.~(\ref{eq:dGds})) are $PS_2/PS_1\simeq0.6,~0.4$, respectively. This suggests that the branching ratios of $\rm{Br}_{2}$ ($\Lambda_b^0\to \bar{D} \Sigma_c K^-$) and $\rm{Br}_{3}$ ($\Lambda_b^0\to \bar{D}^{*} \Sigma_c K^-$) could be the same order as that of
$\rm{Br}_{1}$ ($\Lambda_b^0\to J/\Psi p K^-$). We thus make the \lq data' as $\rm{Br}_{2,3}=\rm{Br}_{1}$, and the uncertainty of $\rm{Br}_{2,3}$ is set to be 10 times as the central values of them. We input these as constraints.

\section{Fitting strategy and pole analysis}
For the K matrix, we try to use as less parameters as possible. Only when more parameters are indispensable to reduce the $\chi^2$ distinctly do we include them. A pole\footnote{We tried to input more poles in the K matrix, but it only helps to improve the fit a little and will not change the conclusion here.} in the K matrix is necessary to fit to the event shape around $P_c(4440)$. Adding a P-wave instead of inputting a K matrix pole is also checked.
It is somehow helpful to distinguish the quantum number of $P_c(4440)$. For the P-wave scattering amplitude we adopt the Blatt-Weisskopf barrier factor representation~\cite{Dai:2014zta}, see the supplement for details.
The following fits are performed:
\begin{itemize}
\item[1)] {\bf Fit~1}: We do not include any poles in the K matrix. $\chi ^2_{{\rm d.o.f}}=1.41$.
\item[2)] {\bf Fit~2}: As in Fit 1 but we include one pole in the K matrix. $\chi ^2_{{\rm d.o.f}}=1.32$.
\item[3)] {\bf Fit~3}: As in Fit 1 we do not include poles in the K matrix, but add a P-wave instead. $\chi ^2_{{\rm d.o.f}}=1.32$.
\end{itemize}
The parameters of all the fits are shown in the supplement.
The fit results are shown in Fig.~\ref{Fig:events}. Our amplitudes fit well to the high statistics LHCb data in 2019~\cite{Aaij:2019vzc}, with $\cos\theta_{P_c}$ weighted. It is worth to point out that this dataset has removed much of the interfering of the $\Lambda^*$, which is in the $K^- p$ channel and most populated at $\cos\theta_{P_c}>0$.  Owing to this our K matrix fit without three body final state interactions is feasible.
The branching ratios of $\rm{Br}_{1}$ is exactly the same as that of PDG, and in most of the fits the $\rm{Br}_{2}$ is of  $10^{-4}$, and  $\rm{Br}_{3}$ is of $10^{-5}$.
Notice that in Fit 1 it does not have the structure around the $\sqrt{s}=4440$~MeV.
\begin{figure}[hpt]
\includegraphics[width=0.235\textwidth,height=0.15\textheight]{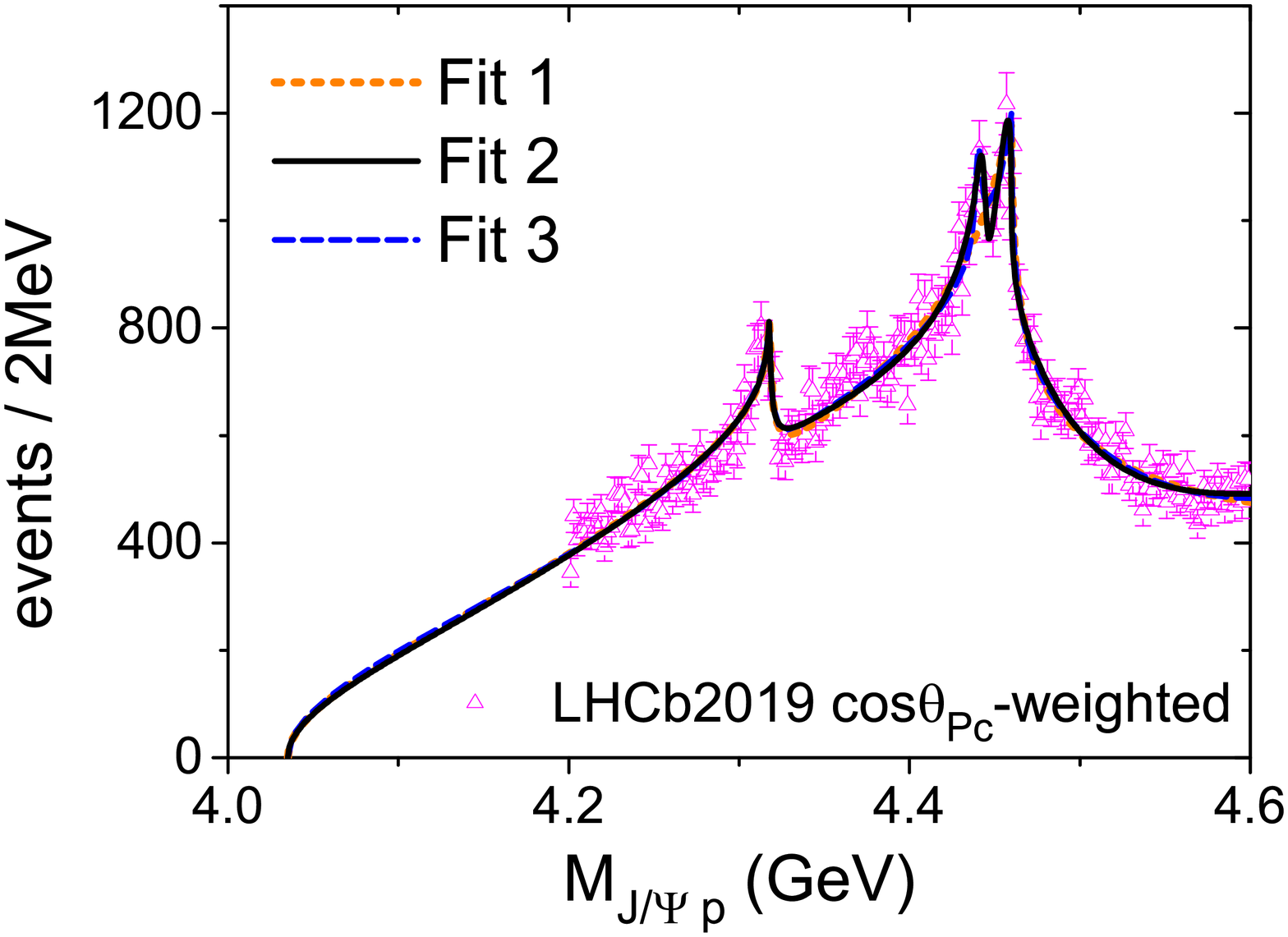}
\includegraphics[width=0.235\textwidth,height=0.15\textheight]{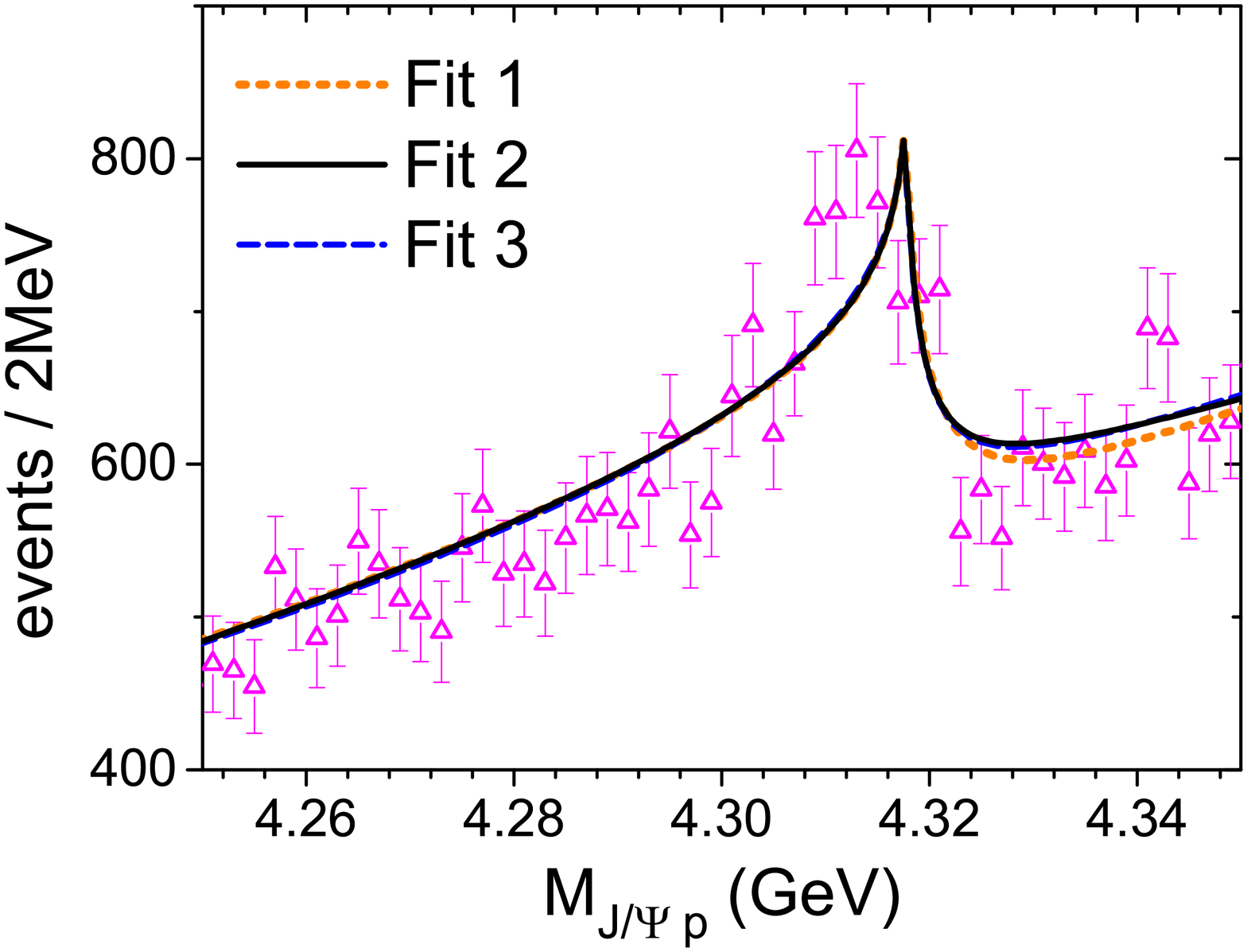}
\caption{\label{Fig:events} Fits to the $J/\Psi p$ spectroscopy of $\Lambda_b\to J/\Psi p K^-$. The right graph is enlarged around the $P_c(4312)$. The LHCb 2019 data is the $\cos\theta_{P_c}$-weighted one from Ref.~\cite{Aaij:2019vzc}.  }
\end{figure}

\begin{figure}[hpt]
\includegraphics[width=0.48\textwidth,height=0.25\textheight]{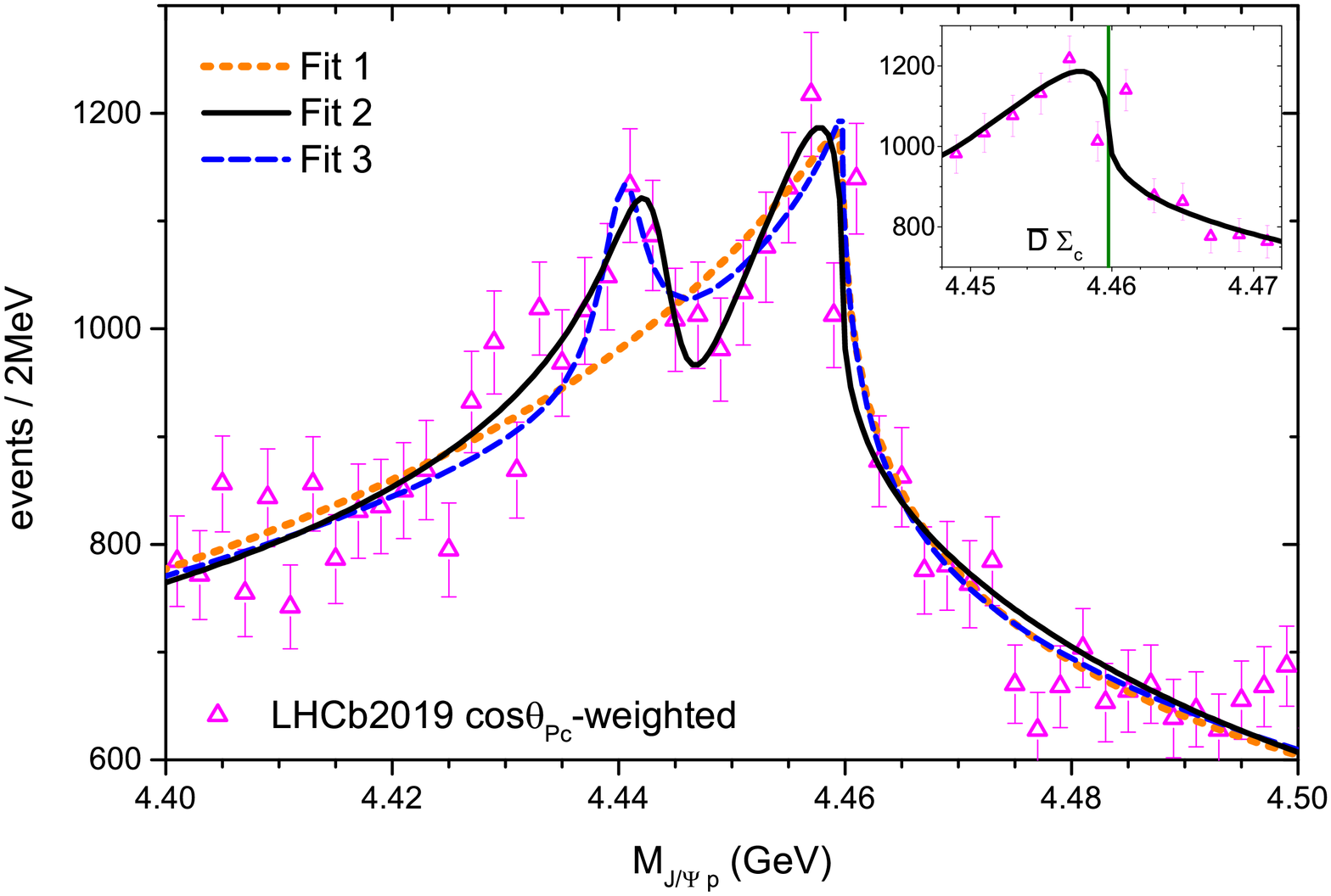}
\caption{\label{Fig:events;2} Fit of the $J/\Psi p$ spectroscopy of $\Lambda_b\to J/\Psi p K^-$ for enlarged size around $P_c(4440)$ and  $P_c(4457)$.
 }
\end{figure}

To study the resonances we enlarge the size of the plots around the structures, as shown in Fig.~\ref{Fig:events;2} and the right side graph of Fig.~\ref{Fig:events}.
For the $P_c(4312)$, our amplitude fits to the data well. Though the amplitude is a bit lower than the data on the left side of the \lq peak', it is within the margin of the data error. Note that we do not input a K-matrix pole around $P_c(4312)$ in the K-matrix formalism. This is why our \lq peak' exactly locates at the $\bar{D}\Sigma_c$ threshold ($\sqrt{s}=4.3177$~GeV), shifting a bit to the right side of the peak of the data.
For the $P_c(4440)$, our Fits 2 and 3 fit to the data well, while in Fit 1 one can not find such a structure.  This is caused by that in Fit 1 we do not include the K matrix pole in $K(s)$.  As shown in Fig.\ref{Fig:events;2}, the Fit 2 is better than Fit 3 in both the $P_c(4440)$ and the $P_c(4457)$ region. Our results prefer an S-wave $P_c(4440)$.
For the $P_c(4457)$, our amplitude behaves more like a cusp caused by the $\bar{D}^{*}\Sigma_c$ threshold effect. These will be discussed in next sections.
We also separate the individual contribution of each channel to the $J/\Psi p$ spectroscopy, that is, we use $F_1=\alpha_i T_{i1}$ to replace of Eq.(\ref{eq:F}). It is found that the $\bar{D}\Sigma_c\to J/\Psi p$ channel dominates the contribution around the $P_c(4312)$, which is consistent with the $\bar{D}\Sigma_c$ molecule picture.
All the three channels contributes like a peak or dip around the $P_c(4440)$, suggesting it to be a Breit-Wigner particle.
Around the $P_c(4457)$ all the channels have non-ignorable structure around $\bar{D}^*\Sigma_c$ threshold, while that of the $\bar{D}^*\Sigma_c \to J/\Psi p$ channel dominates. What is more, the $\bar{D}^*\Sigma_c$ contribution behaves like either a threshold effect or a bound state below the threshold.
These supports that the $P_c(4457)$ could either be caused by cusp effect or a component of $\bar{D}^* \Sigma_c$ molecule. We will discuss it in next sections.

With the amplitudes given by the Chew-Mandelstam formalism, the information of the poles
can be extracted out. We continue the $T(s)$ amplitude to the unphysical Riemann sheets based on
unitarity and analyticity. The definition of the Riemann sheet (RS) could be found in \cite{Krupa:1995fc}. Here we use the following definition \cite{Dai:2012kf} as shown in Table~\ref{tab:continuation}.
\begin{table}[h]
\begin{center}
\begin{tabular}{c c c c c c c c c}
\hline
               & I    & II  & III  & IV  & V    & VI    & VII  & VIII   \\
\hline \hline
$\rho_1$       & $+$  & $-$ & $-$  & $-$ & $+$  &  $+$  & $-$  & $+$      \\
$\rho_2$       & $+$  & $+$ & $-$  & $-$ & $-$  &  $+$  & $+$  & $-$      \\
$\rho_3$       & $+$  & $+$ & $+$  & $-$ & $+$  &  $-$  & $-$  & $-$      \\
\hline
\end{tabular}
\caption{\label{tab:continuation}The sign of phase factors for each Riemann sheet. }
\end{center}
\end{table}
The pole $s_R$ and its coupling/residue of the RS-n in the triple channel are defined as:
\bea
T^{n}_{ij}(s)=\frac{g_{i}g_{j}}{s_R^n-s}\,,\label{eq;g}
\eea
where the subscript \lq $i$, $j$' denote the hadronic channels as before.
The poles in different RSs for all the fits are given in Table~\ref{tab:poles;case}.
\begin{table}[hpt]
{\footnotesize
\begin{tabular}{|c |c | c |c |@{}c @{}|c|@{}c @{}| c |@{}c@{} |}
\hline
\rule[-0.4cm]{0cm}{0.8cm}\multirow{2}{*}{\rule[-0.8cm]{0cm}{1.6cm}State} &  \multicolumn{6}{c|}{ pole locations (MeV)}  \\
\cline{2-7}
\rule[-0.4cm]{0cm}{0.8cm}   & RS  & Fit.1  & RS & Fit.2  & RS & Fit.3   \\
\hline\hline
\multirow{4}{*}{\rule[-2cm]{0cm}{3cm}$P_c(4312)$} & \multirow{2}{*}{\rule[-0.4cm]{0cm}{1.0cm}III}     & $4296.93^{+2.48}_{-3.00}$ &\multirow{2}{*}{\rule[-0.4cm]{0cm}{1.0cm}III} & $\cdots$   & \multirow{2}{*}{\rule[-0.4cm]{0cm}{1.0cm}\textbf{V}$^\star$}    & $4313.38^{+2.52}_{-5.73}$   \rbox \\
&    & $-i5.12^{+2.44}_{-1.06}$  &   & $\cdots$     &     & $-i2.05^{+1.65}_{-0.75}$     \rbox \\[0.5mm]  \cline{2-7}
& \multirow{2}{*}{\rule[-0.4cm]{0cm}{1.0cm}\textbf{V}$^\star$}     & $4312.74^{+1.69}_{-0.67}$ &\multirow{2}{*}{\rule[-0.4cm]{0cm}{1.0cm}\textbf{V}$^\star$} & $4314.31^{+2.06}_{-1.10}$  &\multirow{2}{*}{\rule[-0.4cm]{0cm}{1.0cm}VIII} & $4313.11^{+3.86}_{-4.76}$    \rbox \\
&    &  $-i3.33^{+2.91}_{-1.25}$  &  & $-i1.43^{+1.50}_{-0.57}$   &     & $-i3.11^{+1.63}_{-2.02}$     \rbox \\[0.5mm]
\hline
\multirow{8}{*}{\rule[-2cm]{0cm}{4cm}$P_c(4440)$} & \multirow{2}{*}{\rule[-0.4cm]{0cm}{1.0cm}$\cdots$}   & $\cdots$
& \multirow{2}{*}{\rule[-0.4cm]{0cm}{1.0cm}\textbf{III}$^\star$} & $4444.09^{+2.53}_{-1.48}$  & \multirow{2}{*}{\rule[-0.4cm]{0cm}{1.0cm}\textbf{III}$^\star$} & $4440.53^{+0.47}_{-0.31}$       \rbox \\
&   & $\cdots$ &   & $-i3.10^{+0.53}_{-1.33}$   &   & $-i2.42^{+0.22}_{-0.22}$         \rbox \\[0.5mm]  \cline{2-7}
& \multirow{2}{*}{\rule[-0.4cm]{0cm}{1.0cm}$\cdots$} & $\cdots$  & \multirow{2}{*}{\rule[-0.4cm]{0cm}{1.0cm}IV} & $4443.69^{+2.89}_{-1.34}$   & \multirow{2}{*}{\rule[-0.4cm]{0cm}{1.0cm}IV} & $4440.38^{+0.41}_{-0.19}$                \rbox \\
&   & $\cdots$ &  & $-i0.32^{+1.23}_{-0.04}$    &   & $-i1.40^{+0.59}_{-0.50}$           \rbox \\[0.5mm]  \cline{2-7}
& \multirow{2}{*}{\rule[-0.4cm]{0cm}{1.0cm}$\cdots$} & $\cdots$ & \multirow{2}{*}{\rule[-0.4cm]{0cm}{1.0cm}V} & $4444.22^{+2.72}_{-1.41}$    & \multirow{2}{*}{\rule[-0.4cm]{0cm}{1.0cm}V} & $4440.53^{+0.37}_{-0.30}$          \rbox \\
&   & $\cdots$ &   & $-i2.48^{+0.57}_{-0.67}$     &   & $-i2.32^{+0.27}_{-0.61}$           \rbox \\[0.5mm]  \cline{2-7}
& \multirow{2}{*}{\rule[-0.4cm]{0cm}{1.0cm}$\cdots$}   & $\cdots$ & \multirow{2}{*}{\rule[-0.4cm]{0cm}{1.0cm}VII}  & $4443.84^{+1.93}_{-1.91}$   & \multirow{2}{*}{\rule[-0.4cm]{0cm}{1.0cm}VIII} & $4440.38^{+3.31}_{-0.52}$      \rbox \\
&  & $\cdots$ &  & $-i1.02^{+1.05}_{-0.92}$      &    & $-i1.30^{+4.45}_{-0.50}$        \rbox \\[0.5mm]
\hline
\multirow{4}{*}{\rule[-2cm]{0cm}{3cm}$P_c(4457)$}
& \multirow{2}{*}{\rule[-0.4cm]{0cm}{1.0cm}$\cdots$}    & $\cdots$  & \multirow{2}{*}{\rule[-0.4cm]{0cm}{1.0cm}III} & $4466.53^{+2.13}_{-4.75}$  & \multirow{2}{*}{\rule[-0.4cm]{0cm}{1.0cm}$\cdots$}  & $\cdots$   \rbox \\
&  & $\cdots$ &    & $-i3.88^{+6.95}_{-0.93}$    &   & $\cdots$   \rbox \\[0.5mm]\cline{2-7}
& \multirow{2}{*}{\rule[-0.4cm]{0cm}{1.0cm}$\cdots$}    & $\cdots$ & \multirow{2}{*}{\rule[-0.4cm]{0cm}{1.0cm}VII} & $4456.77^{+3.10}_{-8.89}$  & \multirow{2}{*}{\rule[-0.4cm]{0cm}{1.0cm}VIII} & $4453.44^{+7.11}_{-3.34}$  \rbox \\
&  & $\cdots$ &     & $-i7.77^{+11.07}_{-4.41}$    &     & $-i21.58^{+8.01}_{-6.36}$  \rbox \\[0.5mm]
\hline\hline
\end{tabular}
\caption{\label{tab:poles;case}The pole locations given by our fits. The Riemann sheets with bold type and the \lq$\star$' symbol means that they are close to the physical sheet.  }
}
\end{table}
In Fit 1, we only find poles of $P_c(4312)$. And in Fits 2 and 3 we find poles of all the three resonances.
Fit 2 describes the data better with less assumptions, we choose it as our optimistic one.
To classify the inner structure of the poles, we use the \lq criteria'
proposed in Ref.~\cite{Morgan:1992ge,Dai:2012kf}: A triple channel Breit-Wigner resonance should
appear as quadruplet poles in different RSs, while a molecule has less poles.

\vspace{1mm}
\noindent $\mathbf{P_c(4312)}$\\
This resonance (pole) is rather stable in all the fits. We can find it without an input pole in the $K(s)$.
The poles locate at the RS-III and/or V. For each fit we can find only one or two poles. According to the \lq pole counting', it is a
$\bar{D}\Sigma_c$ molecule.
The masses of the poles are a bit below the threshold $\sqrt{s_{th2}}=4317.73$~MeV,
and their widths (2 times of the imaginary part of the pole) are only a few MeV. This supports the molecule picture.
As is known, RS-II and III are the closest sheet to the physical one below and above $\bar{D}\Sigma_c$ threshold, respectively. RS-II and RS-V are connected along the unitary cut above the $\bar{D}\Sigma_c$ threshold.
The shadow poles appear in RS-V but not in RS-II suggests that there is a strong dynamics to drag
the pole from RS-II to RS-V. The pole in RS-III confirms such observation. This is
consistent with the molecule picture, as our interaction between the resonance and the $\bar{D}\Sigma_c$ is strong and it is typical way how a virtual state  (with weak interaction to the $\bar{D}\Sigma_c$) changes into a molecule. For the strength of the couplings please see the $|g_2|$ of Table~\ref{tab:residue}, in the Appendix A.

\vspace{1mm}
\noindent $\mathbf{P_c(4440)}$\\
The masses of the poles are quite close to the input K matrix pole $4443.60$~MeV in Fit 2.
These poles are found in four sheets: RS-III, IV, V, VII/VIII, being close to each other.
The widths are quite narrow and the poles are quite close to the real axis.
According to \lq pole counting', it is an elementary particle.
Since it decays into $J/\Psi p$ and have five quark component, this implies a compact pentaquark picture.
Note that our widths of the $P_c(4440)$ are small.
Indeed in all our fits except for Fit 1, including Fits A, B and C, all the $P_c(4440)$ have small widths, see also Table~\ref{tab:poles;ABC} in the Appendix.
And all of our fits describe the peak of the data around the Pc(4440) region rather well.
If neither the pole in the K matrix nor the higher partial wave resonance is input, one can not obtain such a peak (see in Fit 1).
Obviously the kinematic behaviour can not supply such a structure. This supports the compact pentaquark picture or it should be a resonance in higher partial waves. However, comparing Fit.2 and 3 in the right side of the $P_c(4440)$, it is obviously to see that the solid black line (Fit.2) fits better to the data than the dashed blue line (Fit.3). This suggests that our fit prefers an S-wave picture of the $P_c(4440)$\footnote{In Refs.\cite{Wu:2010jy,Wu:2010vk,Xiao:2013yca,Liu:2019tjn}, they suggest $P_c(4440)$ to be $1/2^-$, while in Refs.\cite{Yamaguchi:2019seo,Liu:2019zvb}, they suggest  $3/2^-$.  In our case, Fit 3 is worse than Fit 2 and we prefer the $P_c(4440)$ to be S-wave, but it is not possible to distinguish the quantum number.}.

\vspace{1mm}
\noindent $\mathbf{P_c(4457)}$\\
We do not find poles in Fit 1, and find poles in RS-III and VII in Fit 2 and a pole in RS-VIII in Fit 3. For RS-VII and VIII they are faraway from the physical region. Absence of poles close to the physical region means that the structure is caused by cusp effect.
The pole in RS-III (Fit 2) is 7 MeV above the threshold $\sqrt{s_{th3}}=4459.75$~MeV, barely close to the physical region. From the \lq pole counting' it does not support the dominant molecule or \lq Breit-Wigner' types, but a molecular component can not be entirely excluded in Fit 2.
In the region around $\bar{D}^{*} \Sigma_c$, our amplitude behaves more like a cusp but not a normal Breit-Wigner structure.
In Fits 1 and 3 they are caused by cusp effect and in Fit 2 there is a sharp decline near the $\bar{D}^{*} \Sigma_c$ threshold, see the graph in the top-right corner of Fig.\ref{Fig:events;2}. Indeed, this structure is very similar to that of the $\eta'\pi^+\pi^-$ line shape around $\bar{p}p$ threshold, see Fig.4 of \cite{Ablikim:2016itz} and Fig.4 of \cite{Dai:2018tlc}.

\section{Analysis of other datasets}
\begin{figure}[hpt]
\includegraphics[width=0.48\textwidth,height=0.5\textheight]{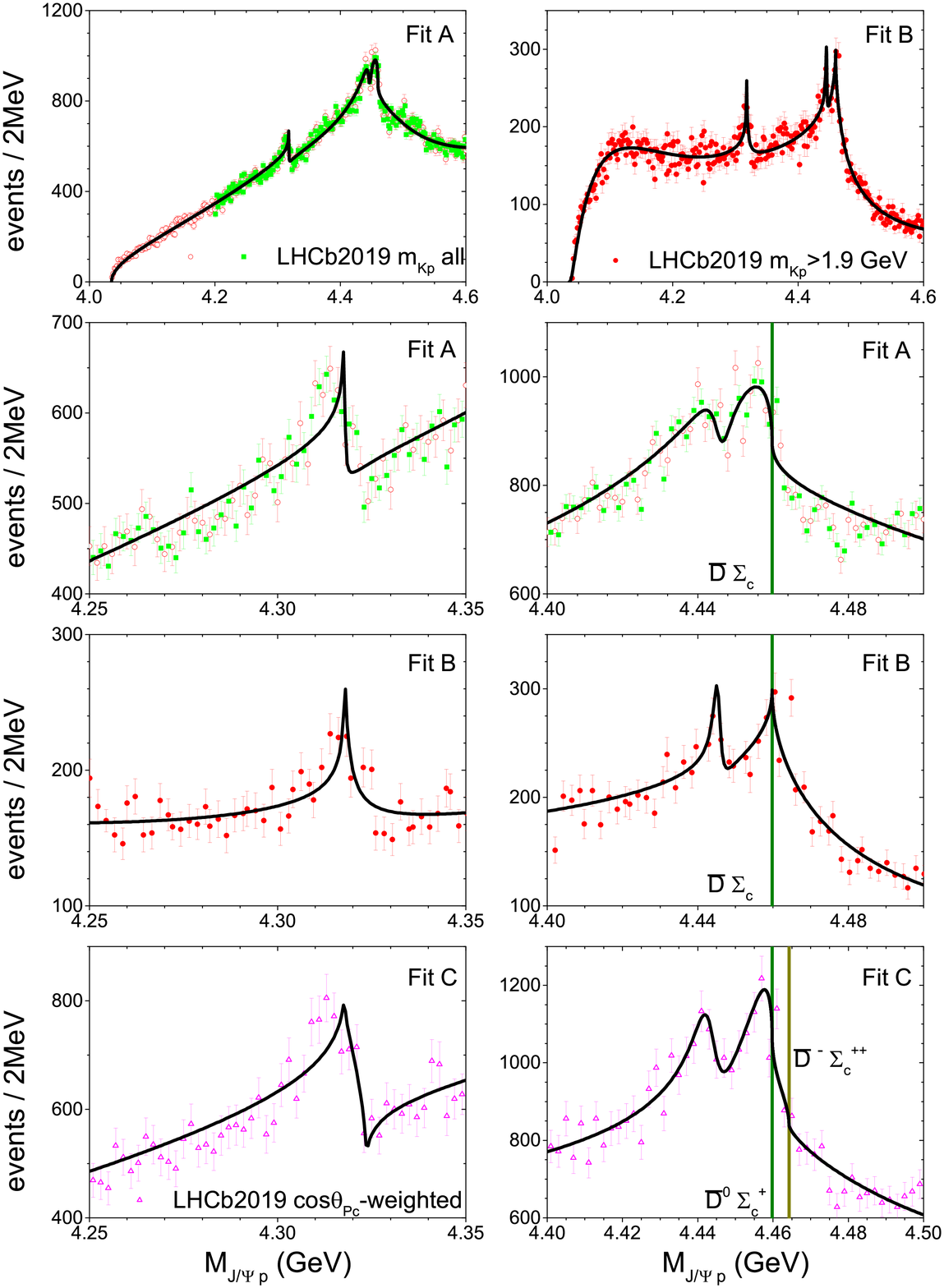}
\caption{\label{Fig:events;ABC} Fit of the $J/\Psi p$ spectroscopy of $\Lambda_b\to J/\Psi p K^-$. All the datasets are from Ref.~\cite{Aaij:2019vzc}. In Fit A the data is from the \lq $m_{Kp}$ all' case, where the red open circle is from Fig.1 and green square is from Fig.3 of Ref.~\cite{Aaij:2019vzc}. In Fit B the data is from the \lq $m_{Kp}>1.9$~GeV' case. In Fit C the data is from the $\cos\theta_{P_c}$-weighted one. The vertical lines denote the $\bar{D}^*\Sigma_c$ thresholds. }
\end{figure}

We also check the case that we fit to other datasets given in Ref.~\cite{Aaij:2019vzc}, the \lq $m_{Kp}$ all' and the \lq $m_{Kp}>1.9$~GeV' ones. Indeed in all these datasets the treatment with the background is different.
Note that the poles are model independent and in all the processes and models their locations should be the same. Thus the poles extracted from different Fits could be used to check the reliability of our analysis.
The following extra fits are performed:
\begin{itemize}
\item[1)] {\bf Fit~A}: As in Fit.2 but we fit to the data of \lq $m_{Kp}$ all' case (without requiring $m_{Kp}>1.9$~GeV).
\item[2)] {\bf Fit~B}: As in Fit.2 but we fit to the data  with requiring $m_{Kp}>1.9$~GeV.
\item[3)] {\bf Fit~C}: As in Fit.2 we fit to the data  with requiring $\cos\theta_{P_c}$ weighted, with the isospin symmetry violation included. That is, the mass difference between $D^{(*)0} \Sigma_c^{+}$  and $D^{(*)-} \Sigma_c^{++}$ is taken into account, see Appendix.A for details.
\end{itemize}
The invariant mass spectrum is given in Fig.\ref{Fig:events;ABC}.
Notice that in Fit C, with isospin violation included, it fits better to the data near thresholds.

The discussion of the event shape structure of these fits are consistent with that given by Fits 1, 2, and 3.
In each fit, there are only one or two poles in RS-III and/or V of the $P_c(4312)$. It supports the molecular component.
And there are four poles of the $P_c(4440)$ in RS-III, IV, V, VII/VIII. They are close to each other and the widths are narrow.
This supports the compact pentaquark component.
For the $P_c(4457)$ we do not find poles nearby in Fits A and B but find two poles in RS-III and VII in Fit C.
Our line shape falls down obviously near the $\bar{D}^*\Sigma_c$ threshold in Fits A and C. And in Fit B ours exhibits a pronounced cusp-like structure around the $\bar{D}^*\Sigma_c$ threshold.
These confirm the cusp effect origin of the $P_c(4457)$, but a component of $\bar{D}^* \Sigma_c$ molecule can't be excluded.
This component could be generated by a virtual state in the  $\bar{D}^*\Sigma_c$ single channel scattering nearby the threshold, see Ref.\cite{Frazer:1964zz} and discussions of the pole trajectory in Appendix B.

\section{Summary}\label{sec:summary}
In this paper we perform an amplitude analysis in the process of $\Lambda_b\to J/\Psi p K^-$. The $J/\Psi p$ ~-~ $\bar{D}^0 \Sigma_c^+$~-~$\bar{D}^{*0} \Sigma_c^+$ triple channel scattering amplitude is constructed by K-matrix, within Chew-Mandelstam formalism.
Based on it we apply the Au-Morgan-Pennington method to study the process of $\Lambda_b\to J/\Psi p K^-$, taking into account the final state interactions.
Qualified fits to the invariant mass spectroscopy \cite{Aaij:2019vzc} is obtained from $J/\Psi p$ threshold up to $\sqrt{s}=4600$~MeV.
We extract out the poles (Fit 2) and find that the $P_c(4312)$, with the pole location $4314.31^{+2.06}_{-1.10}-i1.43^{+1.50}_{-0.57}$~MeV, should have $\bar{D} \Sigma_c$ molecule component. The $P_c(4440)$, $4444.09^{+2.53}_{-1.48}-i3.10^{+0.53}_{-1.33}$~MeV in RS-IV, prefers to be a compact
S-wave pentaquark. The $P_c(4457)$ is most likely to be caused by cusp effect,
while a component of $\bar{D}^* \Sigma_c$ molecule can't be excluded.
We also predict the branching ratios (Fit 2) of the decay of  $\Lambda_b^0\to \bar{D} \Sigma_c K^-$ and $\Lambda_b^0\to \bar{D}^* \Sigma_c K^-$ to be
$(1.49\pm0.26)\times10^{-4}$ and $(0.30\pm0.08)\times10^{-4}$, respectively.
The future LHCb measurement of the decays, for instance $\Lambda_b^0\to D^- \Sigma_c^{++} K^-$ and $\Lambda_b^0\to D^{*-} \Sigma_c^{++} K^-$, will tell us further information about these mysterious resonances.

\section{Acknowledgements}
We are grateful to Profs. Li-Sheng Geng, Feng-Kun Guo and Han-Qing Zheng for their suggestions. Helpful Discussions with Jie-Sheng Yu and Li-Ming Zhang (member of LHCb collaboration) are also acknowledged.
This work is supported by National Natural Science Foundation of China (NSFC) with Grant Nos.11805059, 11805012, 11905258. Joint Large Scale Scientific Facility Funds of the NSFC and Chinese Academy of Sciences (CAS) under Contract No.U1932110, and Fundamental Research Funds for the Central Universities.

\newpage
\appendix
\setcounter{equation}{0}
\setcounter{table}{0}
\renewcommand{\theequation}{\Alph{section}.\arabic{equation}}
\renewcommand{\thetable}{\Alph{section}.\arabic{table}}

\section{K matrix formalism}
In Fits A and B we apply the same method as that of Fit 2.
In Fit C we take into account the contribution of isospin violation, where the phase space factor $\rho_2(s)$ is replaced by $\frac{1}{2}[\rho_{D^- \Sigma_c^{++}}(s)+\rho_{\bar{D}^0 \Sigma_c^{+}}(s) ]$, and $\rho_3(s)$ by $\frac{1}{2}[\rho_{D^{*-} \Sigma_c^{++}}(s)+\rho_{\bar{D}^{*0} \Sigma_c^{+}}(s) ]$. See Ref.\cite{Dai:2014zta} for similar discussions on isospin symmetry breaking in $\overline{K}K$.
\begin{widetext}
The parameters and $\chi^2/d.o.f$ of all our fits are given in Table~\ref{tab:para}.
\begin{table*}[h!]
{\footnotesize
\begin{center}
\begin{tabular}{|c||c|c|c|c|c|c|c|c|}
\hline
                & Fit 1              & Fit 2               & Fit 3              & Fit A               & Fit B              & Fit C        \\
\hline\hline
$s_{1}$(GeV$^2$) & $\cdots$          & $19.7305\pm0.01080$ & $\cdots$           & $19.7320\pm0.0021$ & $19.6418\pm0.0054$ & $19.7305\pm0.01132$     \\
$f^{1}_{1}$(GeV) & $\cdots$          & $0.0383\pm0.0099$   & $\cdots$           & $0.0710\pm0.0065$   & $-0.3539\pm0.0101$ & $0.0462\pm0.0131$     \\
$f^{2}_{1}$(GeV) & $\cdots$          & $-0.0890\pm0.0140$  & $\cdots$           & $-0.0873\pm0.0069$  & $ 0.2660\pm0.0072$ & $-0.0704\pm0.0127$        \\
$f^{3}_{1}$(GeV) & $\cdots$          & $0.0670\pm0.0093$   & $\cdots$           & $0.0801\pm0.0046$   & $0.2434\pm0.0070$  & $0.0583\pm0.0069$           \\
$c^{11}_{0}$    & $-0.5392\pm0.1364$  & $-0.9241\pm0.1060$  & $-0.2112\pm0.0167$ & $-0.8246\pm0.0044$  & $0.2132\pm0.0383$  & $-1.0088\pm0.0857$  \\
$c^{11}_{1}$    & $ 0.1001\pm0.0891$ & $1.3494\pm0.3334$   & $-0.7066\pm0.0399$ & $1.1965\pm0.0110$   & $ 2.7235\pm0.2104$ & $ 1.5565\pm0.3370$ \\
$c^{12}_{0}$    & $0.0396\pm0.0617$ & $ 0.2071\pm0.0761$  & $-0.4497\pm0.908$  & $ 0.2615\pm0.0092$  & $-0.8442\pm0.0009$ & $ 0.1904\pm0.0795$\\
$c^{12}_{1}$    & $-4.7656\pm0.2428$  & $-5.1328\pm0.4706$  & $ 1.1062\pm0.4885$ & $-4.8447\pm0.0794$  & $-1.2281\pm0.0726$ & $-5.5531\pm0.4805$ \\
$c^{22}_{0}$    & $ -0.4085\pm0.0554$  & $-0.3649\pm0.1025$  & $ 1.1611\pm0.2074$ & $-0.3118\pm0.0119$  & $ 2.1457\pm0.0083$ & $-0.5259\pm0.1828$ \\
$c^{22}_{1}$    & $9.9884\pm 0.1046$  & $7.8723\pm0.3215$   & $ 1.9002\pm1.5049$ & $7.1725\pm0.0359$   & $-4.2733\pm0.1219$ & $7.5295\pm0.7121$   \\
$c^{13}_{0}$    & $-0.7807\pm0.2070$ & $-0.0407\pm0.0677$  & $-1.0142\pm0.0505$ & $-0.3276\pm0.0301$  & $-0.7798\pm0.0106$ & $-0.1177\pm0.0430$ \\
$c^{13}_{1}$    & $0.4087\pm0.8399$  & $-1.6132\pm0.2264$  & $ 2.1896\pm0.2172$ & $0.3794\pm0.1622$  & $-1.2419\pm0.0700$ & $-1.0096\pm0.1143$ \\
$c^{23}_{0}$    & $-0.7556\pm0.1362$ & $-1.0730\pm0.0402$  & $-0.3413\pm0.0915$ & $-1.0787\pm0.0026$  & $-0.5174\pm0.0024$ & $-1.1652\pm0.0260$  \\
$c^{23}_{1}$    & $0.2857\pm0.5907$  & $2.6465\pm0.2148$   & $ 2.0355\pm0.4893$ & $2.5469\pm0.0126$   & $3.7056\pm0.0416$  & $3.3315\pm0.1592$ \\
$c^{33}_{0}$    & $1.1530\pm0.1339$  & $1.9358\pm0.1432$   & $-0.7824\pm0.0541$ & $1.9411\pm0.0022$   & $2.8976\pm0.0005$  & $2.4978\pm0.1019$ \\
$c^{33}_{1}$    & $0.4160\pm0.4285$  & $-1.7578\pm0.5785$  & $ 9.2785\pm0.2918$ & $-1.3144\pm0.0359$  & $-3.3015\pm0.0220$ & $-4.0641\pm0.3509$   \\
$10^{6}\alpha_1$ &$3.8552\pm0.1107$  & $4.8465\pm0.1148$   & $ 9.0434\pm0.4427$ & $5.2785\pm0.0305$   & $9.9157\pm0.0239$  & $4.5698\pm0.0713$   \\
$10^{6}\alpha_2$ &$2.7388\pm0.1961$  & $3.1546\pm0.1314$   & $ 1.6652\pm0.0709$ & $2.8896\pm0.0302$   & $9.2770\pm0.1143$  & $3.3063\pm0.0673$  \\
$10^{6}\alpha_3$ &$2.3640\pm0.1966$  & $1.5606\pm0.0743$   & $ 3.2539\pm0.2474$ & $1.4212\pm0.0126$   & $3.0010\pm0.0522$  & $1.2716\pm0.1203$    \\
$10^{-18}N$      &$3.4010\pm0.1292$  & $3.7216\pm0.2288$   & $ 3.9913\pm0.1251$ & $3.8282\pm0.0102$   & $0.7954\pm0.0208$  & $3.6896\pm0.1564$   \\
$M_P$ (GeV)     & $\cdots$           & $\cdots$            & $4.4405\pm0.0006$  & $\cdots$            & $\cdots$           & $\cdots$     \\
$\Gamma_{P\to1}$(MeV)  & $\cdots$    & $\cdots$            & $0.1001\pm0.0170$  & $\cdots$            & $\cdots$           & $\cdots$     \\
$\Gamma_{P\to2}$(MeV)  & $\cdots$    & $\cdots$            & $3.7321\pm1.5880$  & $\cdots$            & $\cdots$           & $\cdots$     \\
$\Gamma_{P\to3}$(MeV)  & $\cdots$    & $\cdots$            & $1.0234\pm0.4089$  & $\cdots$            & $\cdots$           & $\cdots$    \\
$\beta_{P}$            & $\cdots$    & $\cdots$            & $1.1771\pm0.3005$  & $\cdots$            & $\cdots$           & $\cdots$     \\
$\chi ^2_{{\rm d.o.f}}$  & 1.41      & 1.32                & 1.32               & 1.32                &  1.45              &  1.23            \\[1mm]
\hline\end{tabular}
\caption{\label{tab:para} Results of our fits, as explained in the text. The $c^{ij}_0$ is dimensionless and the $c^{ij}_1$ is in unit of GeV$^{-2}$.
The uncertainty of the parameters is given from MINUIT.  }
\end{center}
}
\end{table*}
\end{widetext}

For all the fits, the masses of the resonances are given by the PDG [36],
and the central values are input as: $M_{\Lambda_b^0}=5619.60$~MeV, $M_{J/\Psi}=3096.90$~MeV, $M_p=938.27$~MeV, $M_{\bar{D}^0}=1864.83$~MeV, $M_{\bar{D}^{*0}}=2006.85$~MeV, $M_{\Sigma_c^+}=2452.90$~MeV. We also try to vary the input masses within the uncertainty given by the PDG and find that the fit results change little. In Fit C we also input $M_{\bar{D}^-}=1869.65$~MeV, $M_{\bar{D}^{*-}}=2010.26$~MeV, $M_{\Sigma_c^{++}}=2453.97$~MeV.

To include the P-wave scattering amplitude we adopt the Blatt-Weisskopf barrier factor representation [25]
\be\label{eq:g;P}
\gamma_i^2(s)=\frac{M_{P}~\Gamma_{P\to i}
~\mathcal{Q}_i(M_{p}^2)}{\rho_i(M_P^2)~\mathcal{Q}_i(s)} \;,
\ee
with $\mathcal{Q}_i(s)=1+q^2/(s-(m_i+M_i)^2)$ and $q$ is chosen to be $1$~GeV. Here $M_P$ and $\Gamma_{P\to i}$ are the input mass and width (decaying to the $i$-th channel) of the P-wave resonance in order.
We have the scattering amplitudes
\be\label{eq:T;P}
{T^{P}}_{i}=\frac{\gamma_i(s)^2}{ M^2-s-i \rho_1(s)\gamma_1^2(s)-i \rho_2(s)\gamma_2^2(s)-i \rho_3(s)\gamma_3^2(s) },
\ee
and the relative $\Lambda_b^0$ decay amplitudes
\be\label{eq:F;P}
\mathcal{F}^{P}_{i}=\frac{\beta_P~\gamma_i(s)}{ M^2-s-i \rho_1(s)\gamma_1^2(s)-i \rho_2(s)\gamma_2^2(s)-i \rho_3(s)\gamma_3^2(s) }.
\ee
Note that the partial wave decomposition factor and the coupling $\gamma_{P \Lambda_b^0 K^+}$ is absorbed into the $\beta_P$.

The branching ratios of all the fits are shown in Table~\ref{tab:Br}.
\begin{table}[h!]
{\footnotesize
\begin{center}
\begin{tabular}{|c||c|c|c|c|}
\hline
                 &  $10^4 \rm{Br}_1$  & $10^4 \rm{Br}_2$  &   $10^4 \rm{Br}_3$                \\
\hline\hline
Fit 1 &  $3.20\pm0.27$  & $1.22\pm0.15$  &  $0.84\pm0.19$  \\
Fit 2 &  $3.20\pm0.40$  & $1.49\pm0.26$  &  $0.30\pm0.08$ \\
Fit 3 &  $3.20\pm0.64$  & $2.04\pm0.89$  & $4.82\pm1.20$  \\
Fit A &  $3.21\pm0.31$ & $0.76\pm0.08$   & $0.65\pm0.13$  \\
\textit{Fit B} &  $3.20\pm0.31$ & $9.12\pm0.70$   & $0.63\pm0.06$  \\
Fit C &  $3.20\pm0.36$  & $1.73\pm0.27$   & $0.15\pm0.04$ \\[1mm]
\hline
\end{tabular}
\caption{\label{tab:Br} The $\rm{Br}_{1,2,3}$ denotes the branching ratios of $\Lambda_b^0\to J/\Psi p K^-$, $\Lambda_b^0\to \bar{D} \Sigma_c K^-$, $\Lambda_b^0\to \bar{D}^{*} \Sigma_c K^-$,  respectively.  The PDG [36] value of the branching ratio of $\Lambda_b\to J/\Psi p K^-$  is $(3.2\pm0.6)\times10^{-4}$.  }
\end{center}
}
\end{table}
In all the fits our $\rm{Br}_1(\Lambda_b^0\to J/\Psi p K^-)$ is exactly the same as that of PDG, while in most of fits $\rm{Br}_2$($\Lambda_b^0\to \bar{D} \Sigma_c K^-$) is roughly 1/3 to 2/3 of $\rm{Br}_1$, and the $\rm{Br}_3$($\Lambda_b^0\to \bar{D}^{*} \Sigma_c K^-$) is of the order of $10^{-5}$.
The branching ratio $\rm{Br}_3$ of Fit 3 is much different from those of other fits. Notice that in Fit 3 we include the P-wave. This indicates that measuring the $\rm{Br}_{2,3}$ would be rather important for the amplitude analysis.
Nevertheless, all the $\rm{Br}_{2,3}$ are of the order of $10^{-4}-10^{-5}$.
The uncertainty of the branching ratios is collected by the uncertainty from MINUIT and the statistics of a dozen of other solutions.
We are aware that the amplitude above $\sqrt{s}=4.6$~GeV is not fitted to the data.
However, as we have checked, using a polynomial to fit to the data above $4.6$~GeV and input it in the integration of Eq.~(7), the difference is only several percents, at most 11\%.
We thus use our K matrix amplitude to do the integration in the whole energy region, and input the difference discussed above as part of the uncertainty.
Notice that the branching ratios in Fit B do not have the credibility as our other results.
This is reflected by the italic type.
The reason is that the cut condition $m_{Kp}>1.9$~GeV reduces the event, and the cut out one should also contribute to the branching ratio. In contrast, the $\cos\theta_{P_c}$-weighted data has also cut out the $\Lambda^*$ contibution, but the event shape of it is quite the same as that of the \lq $m_{Kp}$ all' data by multiplying a normalization factor. The latter data does not miss such contribution.

The poles and couplings are shown in Tables~\ref{tab:poles;ABC} and \ref{tab:residue}, respectively.
\begin{table}[hpt]
{\footnotesize
\begin{tabular}{|c |c | c |@{}c @{}| c |@{}c@{} | c |@{}c@{} |}
\hline
\rule[-0.4cm]{0cm}{0.8cm}\multirow{2}{*}{\rule[-0.8cm]{0cm}{1.6cm}State} &  \multicolumn{6}{c|}{ pole locations (MeV)}  \\
\cline{2-7}
\rule[-0.4cm]{0cm}{0.8cm}   & RS & Fit.A    & RS & Fit.B    & RS & Fit.C  \\
\hline\hline
\multirow{2}{*}{\rule[-2cm]{0cm}{2.5cm}$P_c(4312)$} & \multirow{2}{*}{\rule[-0.4cm]{0cm}{1.0cm}$\cdots$}
     & $\cdots$ & $\cdots$ & $\cdots$  & \multirow{2}{*}{\rule[-0.4cm]{0cm}{1.0cm}III}    & $4320.49^{+0.68}_{-1.37}$   \rbox \\
& $\cdots$  & $\cdots$  & $\cdots$    & $\cdots$   &      &  $-i0.38^{+0.31}_{-0.13}$   \rbox \\[0.5mm]  \cline{2-7}
     & \multirow{2}{*}{\rule[-0.4cm]{0cm}{1.0cm}\textbf{V}$^\star$}
& $4317.44^{+0.17}_{-0.17}$ &\multirow{2}{*}{\rule[-0.4cm]{0cm}{1.0cm}\textbf{V}$^\star$} & $4316.50^{+0.02}_{-0.02}$ &\multirow{2}{*}{\rule[-0.4cm]{0cm}{1.0cm}\textbf{V}$^\star$} & $4320.74^{+0.64}_{-4.59}$     \rbox \\
&       &  $-i0.09^{+0.07}_{-0.01}$    &        & $-i0.003^{+0.001}_{-0.001}$ &    & $-i0.66^{+1.34}_{-0.42}$   \rbox \\[0.5mm]
\hline
\multirow{8}{*}{\rule[-2cm]{0cm}{4cm}$P_c(4440)$} & \multirow{2}{*}{\rule[-0.4cm]{0cm}{1.0cm}\textbf{III}$^\star$}
     & $4445.81^{+0.80}_{-4.43}$   & \multirow{2}{*}{\rule[-0.4cm]{0cm}{1.0cm}\textbf{III}$^\star$} & $4445.46^{+3.34}_{-8.78}$   & \multirow{2}{*}{\rule[-0.4cm]{0cm}{1.0cm}\textbf{III}$^\star$} &  $4443.65^{+1.14}_{-1.51}$        \rbox \\
&    & $-i2.77^{+1.09}_{-0.81}$    &           & $-i1.13^{+1.35}_{-0.39}$      &           &  $-i3.50^{+0.69}_{-1.44}$      \rbox \\[0.5mm]  \cline{2-7}
& \multirow{2}{*}{\rule[-0.4cm]{0cm}{1.0cm}IV}
     & $4443.86^{+1.41}_{-1.64}$   & \multirow{2}{*}{\rule[-0.4cm]{0cm}{1.0cm}IV} & $4440.90^{+4.29}_{-1.20}$ & \multirow{2}{*}{\rule[-0.4cm]{0cm}{1.0cm}IV} & $4442.96^{+0.62}_{-2.31}$              \rbox \\
&    & $-i0.35^{+3.44}_{-0.11}$    &           & $-i6.68^{+2.96}_{-2.24}$    &           & $-i0.38^{+3.64}_{-0.07}$         \rbox \\[0.5mm]  \cline{2-7}
& \multirow{2}{*}{\rule[-0.4cm]{0cm}{1.0cm}V}
    & $4445.71^{+0.63}_{-4.04}$     & \multirow{2}{*}{\rule[-0.4cm]{0cm}{1.0cm}V} & $4445.75^{+0.42}_{-1.00}$   & \multirow{2}{*}{\rule[-0.4cm]{0cm}{1.0cm}V} & $4444.06^{+2.08}_{-1.70}$            \rbox \\
&  & $-i2.67^{+0.70}_{-0.80}$      &   &  $-i0.04^{+1.66}_{-0.04}$    &   &   $-i2.81^{+0.75}_{-0.77}$           \rbox \\[0.5mm]  \cline{2-7}
& \multirow{2}{*}{\rule[-0.4cm]{0cm}{1.0cm}VII}
     & $4443.93^{+1.58}_{-0.60}$   & \multirow{2}{*}{\rule[-0.4cm]{0cm}{1.0cm}VIII} & $4441.37^{+4.32}_{-0.47}$    & \multirow{2}{*}{\rule[-0.4cm]{0cm}{1.0cm}VII} & $4442.23^{+3.45}_{-0.79}$   \rbox \\
&  & $-i0.53^{+0.09}_{-0.18}$       &         & $-i10.80^{+3.57}_{-4.37}$    &         &  $-i0.62^{+1.43}_{-0.43}$        \rbox \\[0.5mm]
\hline
\multirow{4}{*}{\rule[-2cm]{0cm}{3cm}$P_c(4457)$}
& \multirow{2}{*}{\rule[-0.4cm]{0cm}{1.0cm}$\cdots$}
    & $\cdots$   & \multirow{2}{*}{\rule[-0.4cm]{0cm}{1.0cm}$\cdots$}  & $\cdots$   & \multirow{2}{*}{\rule[-0.4cm]{0cm}{1.0cm}III}  & $4464.17^{+1.55}_{-6.33}$   \rbox \\
&   & $\cdots$     &   & $\cdots$    &   & $-i5.85^{+6.62}_{-1.92}$    \rbox \\[0.5mm]\cline{2-7}
& \multirow{2}{*}{\rule[-0.4cm]{0cm}{1.0cm}$\cdots$}
    & $\cdots$  & \multirow{2}{*}{\rule[-0.4cm]{0cm}{1.0cm}$\cdots$} & $\cdots$  & \multirow{2}{*}{\rule[-0.4cm]{0cm}{1.0cm}VII} & $4462.56^{+3.07}_{-14.82}$  \rbox \\
&   & $\cdots$    &            & $\cdots$ &            &  $-i3.90^{+3.13}_{-3.64}$  \rbox \\[0.5mm]
\hline\hline
\end{tabular}
\caption{\label{tab:poles;ABC}The pole locations and residues given by our fits. The Riemann sheets with bold type and the \lq$\star$' symbol means that they are close to the physical sheet. For each pole the locations of all the fits on these sheets have difference at the order of 0.1-1~MeV.  }
}
\end{table}
\begin{table}[hpt]
{\footnotesize
\begin{tabular}{|c |c | c |c |c |c |}
\hline
\rule[-0.3cm]{0cm}{0.7cm} State & Fit  & $|g_{1}|$   & $|g_{2}|$ & $|g_{3}|$  \\
\hline\hline
\multirow{3}{*}{\rule[-1cm]{0cm}{2.8cm}$P_c(4312)$}
& Fit.1~(RS-V)  & $0.29^{+0.03}_{-0.05}$ & $2.17^{+0.64}_{-0.66}$   & $1.92^{+0.52}_{-0.57}$      \rbox \\[0.5mm]  \cline{2-5}
& Fit.2~(RS-V)  & $0.20^{+0.43}_{-0.04}$ & $1.51^{+0.42}_{-0.42}$   & $2.29^{+1.17}_{-0.64}$      \rbox \\[0.5mm]  \cline{2-5}
& Fit.3~(RS-V)  & $0.24^{+0.24}_{-0.19}$ & $1.35^{+0.21}_{-0.20}$   & $0.09^{+0.20}_{-0.09}$      \rbox \\[0.5mm]   \cline{2-5}
& Fit.A~(RS-V)    & $0.05^{+0.01}_{-0.02}$   & $0.70^{+0.26}_{-0.09}$   & $1.19^{+0.19}_{-0.19}$      \rbox \\[0.5mm]  \cline{2-5}
& Fit.B~(RS-V)    & $0.01^{+0.01}_{-0.01}$   & $0.86^{+0.01}_{-0.01}$   & $1.43^{+0.02}_{-0.02}$      \rbox \\[0.5mm]  \cline{2-5}
& Fit.C~(RS-III)  & $0.62^{+0.32}_{-0.21}$ & $1.79^{+1.41}_{-0.95}$   & $2.58^{+3.42}_{-2.58}$      \rbox \\[0.5mm]  \cline{2-5}
& Fit.C~(RS-V)  & $0.59^{+0.16}_{-0.12}$ & $1.70^{+1.10}_{-0.30}$   & $2.45^{+5.31}_{-2.45}$      \rbox \\[0.5mm]  \hline
\multirow{3}{*}{\rule[-1cm]{0cm}{2.2cm}$P_c(4440)$}
& Fit.2~(RS-III)  & $0.09^{+0.02}_{-0.03}$               & $0.34^{+0.15}_{-0.04}$   & $0.51^{+0.13}_{-0.12}$      \rbox \\[0.5mm]  \cline{2-5}
& Fit.3~(RS-III)  & $0.03^{+0.04}_{-0.01}$               & $0.27^{+0.05}_{-0.05}$   & $0.22^{+0.08}_{-0.07}$      \rbox \\[0.5mm]   \cline{2-5}
& Fit.A~(RS-III)  & $0.43^{+0.02}_{-0.02}$   & $0.83^{+0.13}_{-0.12}$   & $1.70^{+0.24}_{-0.21}$      \rbox \\[0.5mm]  \cline{2-5}
& Fit.B~(RS-III)  & $0.12^{+0.01}_{-0.01}$   & $0.15^{+0.13}_{-0.03}$   & $0.77^{+0.16}_{-0.18}$       \rbox \\[0.5mm]   \cline{2-5}
& Fit.C~(RS-III)  & $0.10^{+0.05}_{-0.01}$               & $0.37^{+0.15}_{-0.09}$   & $0.70^{+0.21}_{-0.14}$      \rbox \\[0.5mm]
\hline
\multirow{3}{*}{\rule[-1cm]{0cm}{1.8cm}$P_c(4457)$}
& Fit.2~(RS-III)  & $0.39^{+0.32}_{-0.07}$               & $0.51^{+0.76}_{-0.33}$   & $1.48^{+0.89}_{-0.04}$      \rbox \\[0.5mm]  \cline{2-5}
\multirow{3}{*}{\rule[-1cm]{0cm}{0.5cm}}
& Fit.3~(RS-VIII) & $0.69^{+0.32}_{-0.05}$               & $0.65^{+0.79}_{-0.65}$   & $2.12^{+0.24}_{-0.52}$      \rbox \\[0.5mm]   \cline{2-5}
& Fit.C~(RS-III)  & $0.35^{+0.05}_{-0.02}$               & $0.51^{+1.10}_{-0.15}$   & $1.66^{+2.69}_{-1.00}$      \rbox \\[0.5mm]
\hline\hline
\end{tabular}
\caption{\label{tab:residue}The residues of each poles given by our fits. The unit is GeV.
We only show the one is closest to the physical region. The notation \lq 1, 2 and 3' denotes the channel that the pole couples to.  }
}
\end{table}
For the $P_c(4312)$, we only find one pole in RS-V of Fits A and B. 
Roughly, in each Fit the residue coupling to $\bar{D}\Sigma_c$ ($g_2$) is much larger than that coupling to $J/\Psi p$ ($g_1$).
This supports the \lq $\bar{D}\Sigma_c$' molecule picture, being in compatible with that given by the \lq pole counting' rule.  While it
has more or less the same order magnitude as that coupling to $\bar{D}^*\Sigma_c$ ($g_3$). It reveals the strong transition between the $\bar{D}\Sigma_c$
and $\bar{D}^*\Sigma_c$ channels.
For the $P_c(4440)$, we find poles in four RSs for all the fits, the same as that of Fits 1, 2 and 3, which fits to the $\cos\theta_{P_c}$-weighted dataset.
As discussed above, it is most likely to be a compact pentaquark. The residue $g_{3}(P_c(4440))$, except for Fits 3 and A, is smaller than $g_{3}(P_c(4312))$ and  $g_{3}(P_c(4457))$. This supports that the $P_c(4440)$ is not $\bar{D}^*\Sigma_c$ molecule origin.
For the $P_c(4457)$, we do not find poles close to the 'strcuture' as indicated by the data in Fits A and B, while in Fit C the poles are barely close to the physical RS.  Indeed, the event shapes of Fit A and C are similar to that of Fit 2, dropping rapidly around the $\bar{D}^*\Sigma_c$ threshold. And in Fit B the line shape has a very clear cusp around the threshold.
It confirms that the $P_c(4457)$ is caused by cusp effect at the $\bar{D}^*\Sigma_c$ threshold.
However, a component of $\bar{D}^*\Sigma_c$ molecule can't be excluded, see the $g_3(P_c(4457))$ of Fits 2, 3 and C.
It is much larger than $g_{1,2}(P_c(4457))$, though the poles are barely close to the physical RS.

\section{Pole trajectory}
\begin{figure}[hpt]
\includegraphics[width=0.48\textwidth,height=0.6\textheight]{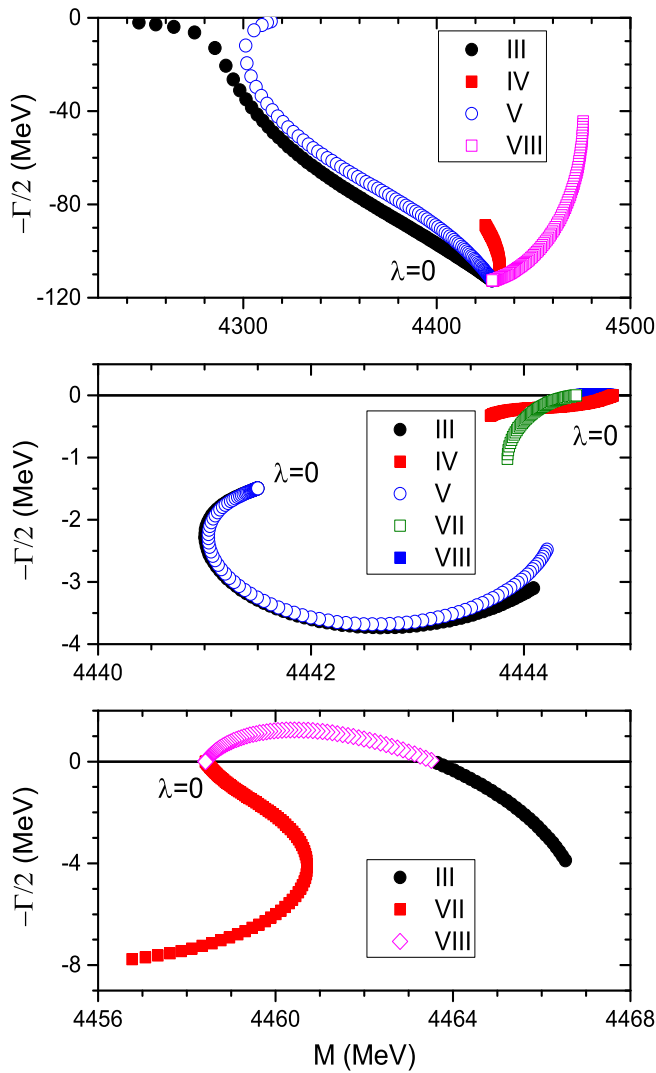}
\caption{\label{Fig:pole;traj} The trajectories of the poles of Fit 2. $\lambda$ is varied from 1 to 0, and the step of $\Delta \lambda$ is 0.01. }
\end{figure}

In this section we track the trajectories of poles by reducing the magnitude of inelastic channels. That is, we change $K_{ij}(s)\rightarrow\lambda K_{ij}(s)$, where $i\neq j$ and vary $\lambda$ from one to zero.
The trajectories of all the poles of Fit 2 are shown in Fig.\ref{Fig:pole;traj}.
It should be noticed that the amplitude at $\lambda=0$ is not physical and the pole could run into weird position. However, as discussed below, it can still give useful hints about the resonances.

\vspace{1mm}
\noindent $\mathbf{P_c(4312)}$\\
\noindent  Except for the pole in RS-V, there are three other shadow poles in RS-III, IV and VII. These shadow poles are dragged far away from the pole of RS-V, due to the strong interaction between the pole and the $\bar{D}\Sigma_c$ channel. Finally all of them merge into one \lq destination pole', which is a resonance ($\sqrt{s_p}=4428.56 - i 112.57$~MeV) in RS-II of $\bar{D}\Sigma$ single channel scattering.
Since all the poles in different RSs are originated from $\bar{D}\Sigma$ scattering, it supports the $\bar{D}\Sigma$ molecular picture. This conclusion is
consistent with the \lq pole counting' rule.

\vspace{1mm}
\noindent $\mathbf{P_c(4440)}$\\
There are four poles nearby $\sqrt{s}=4440$~MeV in RS-III, IV, V, VII, supporting the non-molecular origin.
The poles in RS-III and V merge together at the destination pole ($\sqrt{s_p}=4441.50 - i1.49$~MeV), a resonance in RS-II of $\bar{D}\Sigma_c$ single channel scattering.
And the poles in RS-IV and VII merge together at the destination pole in the real axis ($\sqrt{s_p}=4444.83$~MeV), a virtual state in RS-II of $\bar{D}^*\Sigma_c$ single channel scattering.
Unlike a molecule which couples strongly to the single channel $\bar{D}\Sigma_c$, an elementary particle could couples to multi-channels strongly. See the couplings shown in Table~\ref{tab:residue} for the $P_c(4440)$.
When the inelastic channels are shut down, the coupled channel scattering is spitted into several single channel scattering,  and the destination pole in different single channel behaves differently. The trajectory supports the elementary particle picture (compact pentaquark).

\vspace{1mm}
\noindent $\mathbf{P_c(4457)}$\\
Only in Fits 2, 3 and C do we find poles close to the $P_c(4457)$ region, and all these poles are not close to the physical sheet.
No pole close to the physical region suggests that the structure is generated by the cusp effect.
The $P_c(4457)$  pole trajectory of RS-III will go across the cut and get into RS-VIII, finally it meets the pole coming from RS-VII at the real axis.
The destination pole ($\sqrt{s_p}=4458.42$~MeV) lies closely below the $\bar{D}^* \Sigma_c$ threshold ($\sqrt{s_{th3}}=4459.75$~MeV) and it is the virtual state in RS-II of the $\bar{D}^* \Sigma_c$ single channel. As discussed above, a virtual state close to threshold could also generates \lq cusp' structure around threshold, see in Ref.[43]. This means that a component of $\bar{D}^* \Sigma_c$ molecule is also possible, and it could be coming from a virtual state origin in single channel.
However, our pole analysis reveals that most likely the threshold effect generates the structure of the $P_c(4457)$.


\end{document}